\documentclass[conference]{IEEEtran}
\IEEEoverridecommandlockouts
\usepackage{cite}
\usepackage{amsmath,amssymb,amsfonts}
\usepackage{algorithmic}
\usepackage{graphicx}
\usepackage{textcomp}
\usepackage{xcolor}
\usepackage{url}
\usepackage[pangram]{blindtext}
\usepackage{booktabs}
\usepackage[normalem]{ulem}
\usepackage{ulem}

\def\BibTeX{{\rm B\kern-.05em{\sc i\kern-.025em b}\kern-.08em
    T\kern-.1667em\lower.7ex\hbox{E}\kern-.125emX}}

\makeatletter
\def\endthebibliography{
  \def\@noitemerr{\@latex@warning{Empty `thebibliography' environment}}%
  \endlist
}

 \let\old@ps@headings\ps@headings
 \let\old@ps@IEEEtitlepagestyle\ps@IEEEtitlepagestyle
 \def\confheader#1{%
 \def\ps@headings{%
 \old@ps@headings%
 \def\@oddhead{\strut\hfill#1\hfill\strut}%
 \def\@evenhead{\strut\hfill#1\hfill\strut}%
 }%
 \def\ps@IEEEtitlepagestyle{%
 \old@ps@IEEEtitlepagestyle%
 \def\@oddhead{\strut\hfill#1\hfill\strut}%
 \def\@evenhead{\strut\hfill#1\hfill\strut}%
 }%
 \ps@headings%
 }
 \makeatother
 
\begin{document}

\title{Business-Driven Technical Debt Prioritization:\\An Industrial Case Study
\thanks{This work is partially supported by INES, CNPq grant 465614/2014-0 and CAPES grant 88887.136410/2017-00.}
}

\author{
    \IEEEauthorblockN{Rodrigo Rebouças de Almeida\IEEEauthorrefmark{1}\IEEEauthorrefmark{3}, Rafael do Nascimento Ribeiro\IEEEauthorrefmark{3},  Christoph Treude\IEEEauthorrefmark{2}, Uir\'{a} Kulesza\IEEEauthorrefmark{3}}
\IEEEauthorblockA{\IEEEauthorrefmark{1}Federal University of Paraíba - UFPB, Rio Tinto, PB, Brazil}
\IEEEauthorblockA{\IEEEauthorrefmark{2}University of Adelaide, Adelaide, Australia}
\IEEEauthorblockA{\IEEEauthorrefmark{3}Federal University of Rio Grande do Norte - UFRN, Natal, RN, Brazil}
rodrigor@dcx.ufpb.br, rafaelribeiro@ufrn.edu.br, christoph.treude@adelaide.edu.au,  uira@dimap.ufrn.br}

\maketitle

\begin{abstract}

Incorporating the business perspective into prioritizing technical debt is essential to contribute to decision making in industry. In this paper, we evolve and evaluate a business-driven approach for technical debt prioritization. The approach was evaluated during a five-months industrial case study with business and technical stakeholders' active participation. The results show that the approach contributed to aligning business criteria between the business and technical stakeholders.  We also observed a downward trend in the amount of technical debt that affects high-value business assets. Moreover, we identified eight business factors that affect the decision making related to the prioritization of technical debt. The study results suggest that the proposed business-driven technical debt prioritization approach can help teams to focus their efforts on paying off the business' most relevant debt.
\end{abstract}

\section{Introduction}

Technical debt is a metaphor for describing a design or implementation construct that is expedient in the short term, but that sets up a technical context that can make a future change more costly or impossible \cite{tdmbook}.

Causes related to planning and management are protagonists among those responsible for creating technical debt. For example, tight schedules, competitiveness, changes in business prioritization, and business dynamics are responsible for creating a turbulent environment that leads to technical debt \cite{rios2019}.

There is a plethora of work addressing technical debt at different levels. However, the field still lacks a proper treatment considering business aspects \cite{AMPATZOGLOU:2015,avgeriou_et_al:2016,slr-prioritization-2020,RIOS2018-tertiary}. Regarding technical debt prioritization, it is common to find that the criteria, tools, and approaches used to prioritize technical debt lack a business perspective. Lenarduzzi et al.~\cite{terese2019} conducted a systematic literature review on technical debt prioritization and identified only three papers \cite{martini-boch-2015, YLIHUUMO2016, gupta2016} that use business-related constraints. 

Typically, a development team has a decision-making chain that involves different stakeholders, e.g., a technical leader coordinates the development team and negotiates with the product owner (PO), who negotiates with and reports to a higher-level business stakeholder. In this context, the lack of alignment of opinions regarding the business priorities hinders the technical debt prioritization \cite{icsme2018}. According to Mavengere et al. \cite{mavengere2020}, the business-IT alignment is still a challenge for business. The authors point to human tensions and knowledge silos as misaligned factors, among others.

In this paper, we report the results of a case study to evaluate an approach from our previous work~\cite{icsme2019} to support the technical debt prioritization from a business perspective. It contributes to the alignment between technical and business perspectives for technical debt prioritization. 

Following this work, we put our approach\cite{icsme2019} to the ground to evaluate real scenarios and decision-making. The case study took place at Phoebus Technology,\footnote{\url{http://www.phoebus.com.br/}, \url{https://www.paystore.com.br/en}} 
a company which currently provides electronic payment, credit card processing, and sales processing solutions for more than 90 customers including supermarket chains and credit card network stakeholders, e.g., credit card processors, banks, acquirers, and merchants. 

Business-driven technical debt prioritization involves many aspects including information about IT artifacts, different stakeholders, their perspectives, and decision making. To investigate how business decisions affect technical debt prioritization, we ran an industrial case study to answer the following research questions:

\textbf{RQ1: How does the proposed business-driven approach impact technical debt prioritization?}

To answer this question, we applied the proposed approach supported by a tool in a five-months case study where we associated technical debt items to business-value elements; we identified and solved conflicting business perspectives among stakeholders; and, finally, we observed a higher downward trend in the amount of technical debt that  has high business priority.

\textbf{RQ2: What are the business stakeholders' perceptions regarding factors that influence technical debt prioritization?}

By answering RQ2, we found that much goes on behind the scenes regarding the prioritization decision making. After a set of interviews and focus groups, we identified eight business factors that affected the stakeholders' decision making.

\section{Business-driven Technical Debt Prioritization}

\begin{figure}
\begin{center}
\caption{Tracy framework}
\includegraphics[width=.9\columnwidth]{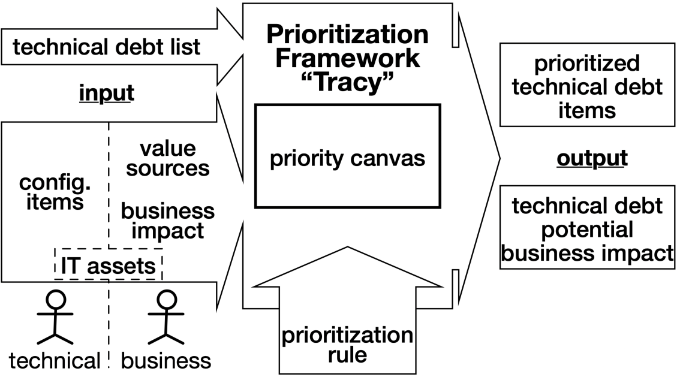}
\vspace{-.9cm}
\label{framework}
\end{center}
\end{figure}

Figure \ref{framework} describes the elements of the business-driven technical debt prioritization framework, called Tracy~\cite{icsme2019}. The approach is constructed around the ``priority canvas,'' detailed in the next section. Besides a technical debt set to be prioritized (``technical debt list''), the Tracy framework uses  configuration items and IT Assets to identify the affected business aspects. The value sources and business impact are the elements that represent the business value. The prioritization rule defines how the technical debt items must be prioritized. Finally, the priority canvas guides the output ``prioritized technical debt items,'' and the business impact canvas guides the other output ``technical debt business impact''. All elements are detailed in the following.

\subsubsection{Technical Debt List}
The technical debt list can be registered directly in the TDM tool or imported from an existing issue tracker system.

\subsubsection{Configuration Item} A term used by SWEBOK~\cite{swebok} and ITSM~\cite{itil} to refer to a managed artifact. In our context, a configuration item (CI) is an artifact that can be affected by technical debt, such as a code fragment, a class, a library, a module, a system, a database, a server, an architecture element, or a service. Configuration items are generally from the technical stakeholders' domain whereas business stakeholders often do not master information about CIs and their responsibilities. A CI can be composed of and dependent on other CIs.

In our model, a configuration item can have one of three states: 
\begin{itemize}
  \item \textbf{operational (oper)}: an artifact that is in production, being directly or indirectly used by customers or users;
  \item \textbf{to-be operational (to-be)}:  a new artifact that is under development or under planning;
  \item \textbf{legacy}: an artifact that is planned to be discontinued or replaced by another one.
\end{itemize}

A technical debt item affects one CI. For example, a ``test debt item'' can affect a ``module of service''; a ``build debt item'' can affect an application; and an ``architectural debt item'' can involve server instances. Configuration items support IT Assets.

\subsubsection{IT Asset} is an abstract concept representing any solution, product, service, or mobile app that is part of the company portfolio. IT Assets are supported by configuration items. This concept's objective is to detach the company's solutions from the technical elements that implement them.  Both technical and business stakeholders usually understand this concept. For example, at the business level, an e-commerce software system is evaluated independent of its technical implementation. It can even share configuration items with other IT assets, e.g., white-label solutions.

IT Assets have the same (but independent) states as configuration items. For example, a \textit{to-be operational} IT asset can be composed of operational CIs (e.g., in the case of a new planned system that uses existing systems). On the other hand, an ``operational'' IT asset must be composed of at least one ``operational'' CI. For example, a new ``to-be'' e-commerce mobile application planned to be released in a few months can be composed of ``to-be'' CIs, e.g., iOS and Android mobile apps, and operational microservices.

\subsubsection{Value Source} In the previous version of this approach \cite{icsme2018, icsme2019}, they used the concept of ``business process'' \cite{bpmbook} to identify the business value of a system. Although every system's features somehow affect a business process, the concept is not ``common sense'' among stakeholders. Both business and technical stakeholders have a different understanding of what a business process is. Sometimes, the business processes or activities are perceived as the ``customer journey'' from the marketing and UX perspective \cite{customerjourney,userjourney}. Besides that, some key features create value, for example, an executive report that supports decision making. Thus we called this abstract concept a Value Source. Value Sources are everything that creates business value from an IT Asset.

A value source can be classified as \textit{core}, when it is part of the core business of a system, or \textit{other}, when it is not. For example, in an e-commerce solution, the features supporting the customer's buying experience are the core business. On the other hand, the ``management of past purchases'' feature is not core business.

The value source can also be classified regarding its \textit{usage frequency}. The \textit{usage frequency} defines how frequently a value source is being used. This information can be obtained by monitoring tools or based on the stakeholders' perception.

Note that to help understanding, we will refer to the relationships between IT Assets or configuration items and value sources as follows: an \textit{oper/core/high} IT Asset refers to an operational IT Asset related to a \textit{core} value source that has \textit{high} usage. Similarly, a \textit{to-be/other/low} IT Asset refers to a not-yet operational IT Asset that affects a value source that is not core-business and has low usage frequency.

\subsection{Priority Canvas}

The Priority Canvas is a board used to visualize the main entities involved in technical debt prioritization. 
This board is used to help stakeholders visualize and discuss IT assets and value sources and their relationships. The board's objective is to guide the participants with exercises to think about IT Assets, value sources, and their classification. The participants can look at ``the same page'' and discuss business-value perceptions. 

\begin{figure}
\begin{center}
\caption{Priority canvas}
\includegraphics[width=1\columnwidth]{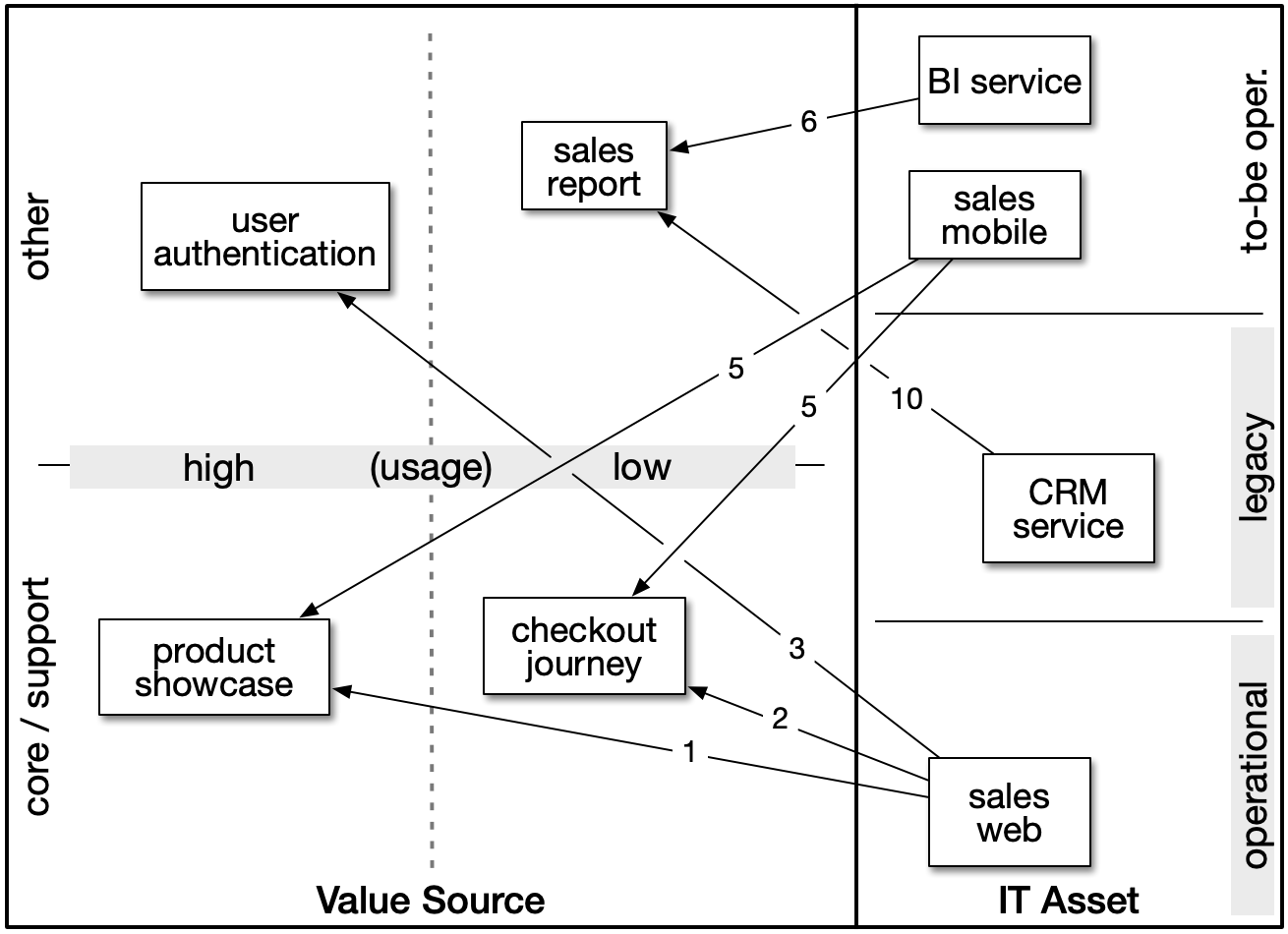}
\vspace{-.5cm}
\label{fig_priority_canvas}
\end{center}
\end{figure}

\begin{table}
\begin{center}
\caption{Priority Rule (example)}
\includegraphics[width=.6\columnwidth]{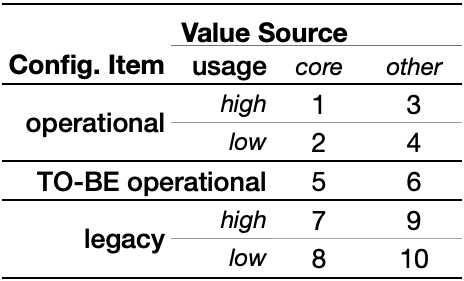}
\vspace{-.5cm}
\label{table_priorityRule-example}
\end{center}
\end{table}

Figure \ref{fig_priority_canvas} shows the board with an illustrative example where the \textit{operational}, \textit{legacy}, and \textit{to-be operational} IT Assets support value sources classified according to their business value (\textit{core/support} and \textit{other}) and their usage frequency (\textit{high} or \textit{low}). The Value Source's \textit{usage} attribute can only be related to operational and legacy IT Assets. 
Since a \textit{to-be} IT Asset is not yet being used, we do not consider its usage frequency for the technical debt prioritization.

\subsubsection{Priority rule}
A priority rule is based on the relationship between IT Assets and value sources. A priority rule classifies the technical debt business priority from 1 (highest priority) to 10 (lowest priority), assigning the relationships between IT Assets and value sources. For example, if we assign a priority 1 to the \textit{oper/core/high} relation, we consider that a technical debt item that affects a highly used core-business value source of an operational asset will have the highest priority. On the other hand, a priority 10 assigned to \textit{legacy/other/low} will set the lowest priority to technical debt that affects non-core-business value sources with low usage supported by legacy IT assets.

Table \ref{table_priorityRule-example} shows the example priorities used in the canvas (Figure \ref{fig_priority_canvas}). In the example, the relation between operational IT assets (sales web) and the highly used core value source (product showcase) has the highest priority 1. Since ``sales mobile'' is a \textit{to-be operational} IT Asset, the priority between ``sales mobile'' and \textit{high} and \textit{low}-use \textit{core} value sources are identical: 5.

Note that the rules are flexible. It is possible to assign any scale of numbers and group different relations with the same priority. For example, in our study (Table~\ref{table_prioritizations}), product owners P6 and P7 suggested grouping technical debt in priorities 2 and 3, respectively. In Section~\ref{sec_percep_prior}, we observe that many factors may affect the prioritization rule. In the study, we also mapped the priorities to the \textit{high-medium-low} scale so we could compare them with the technical priority: priorities 1 to 3 were considered \textit{high}, 4 to 6 \textit{medium}, 7 to 9 \textit{low}, and 10 \textit{lowest}.

\section{Case study}

\subsection{Case study environment}

To answer the research questions, we ran a case study during five months with a team from Phoebus Technology, a company that provides systems for the electronic payment market. The company uses an agile development process with the team's structure inspired by Spotify's Squad model~\cite{spotify}. 

The teams (or squads) are composed of multidisciplinary roles and are responsible for a set of products and services. Each squad has a product owner (PO), who works as a business analyst responsible for concerns such as customer contact, feature specification and prioritization, and delivery planning. The PO reports directly to the CEO. Consisting of developers, testers, software architects, and other roles, the squad is led by the Squad leader, a senior technical professional responsible for managing the squad's development routine and ceremonies, such as sprint planning and creation, and assignment and monitoring of development tasks.

\subsubsection{Selected case} The squad in which the case study took place comprises the PO, the squad leader, six developers, two testers, and an architect. The team is responsible for maintaining eight products, services, and mobile applications that address the business of seven corporate customers. The systems comprise sales and other transaction solutions, business intelligence, integration solutions, and mobile applications that serve supermarket chains and sub-acquires. 

The team uses Redmine\footnote{\url{https://www.redmine.org/}}  as the issue tracking system to maintain a backlog of requirements, development activities, bugs, and other activities related to the development process. They did not manage technical debt in a structured way before this case study. They are subjected to high business pressure, frequent changes in the prioritization of features, and creating new products and features to make sales presentations. 

Unlike the other company's squads, the selected squad has a low dependency on other teams and other products maintained by the company. It is affected by constantly changing business priorities, since they are responsible for a company's new business area, attending to new customers. The product owner also has an active role in the specification of products and services features, prioritizing them, and planning the system's deliveries. 

The company provided us with access to collect data and participate as an observer in several meetings and ceremonies. We as researchers also had permission to schedule meetings with all participants, and we were able to frequently access the top business level, like the CEO (Chief Enterprise Officer) and CSO (Chief Strategic Officer).

\subsection{Case study protocol}

\begin{figure*}[ht]
\begin{center}
\caption{Case study protocol}
\centerline{\includegraphics[width=.8\textwidth]{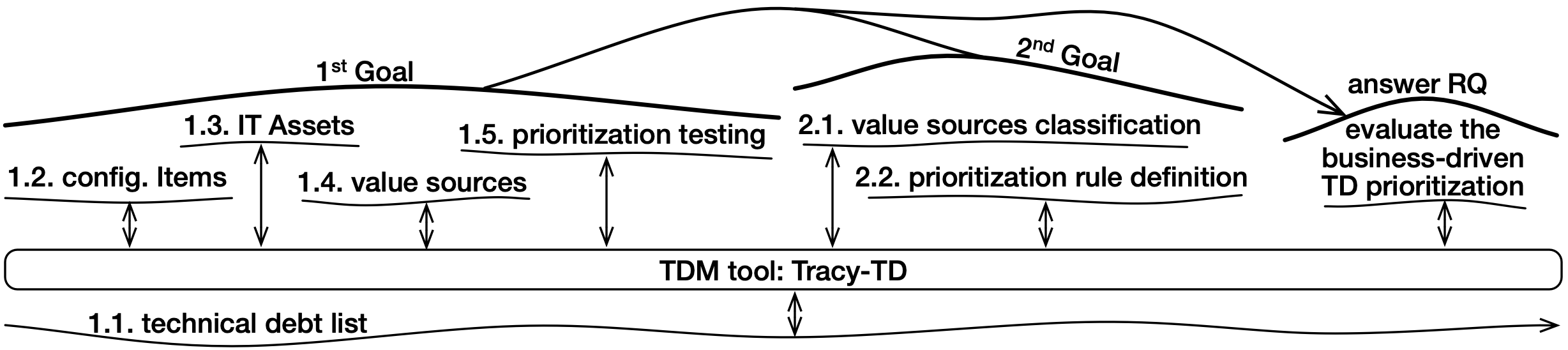}}
\vspace{-0.8cm}
\label{fig_steps}
\end{center}
\end{figure*}

To answer RQ1 (How does the proposed business-driven approach impact the technical debt prioritization?), first, we collected data and prioritized technical debt (1st goal below); and after (Figure~\ref{fig_steps}), we classified the value sources and defined a prioritization rule (2nd goal below) to run and evaluate the business-driven technical debt prioritization.

To answer RQ2 (What are the business stakeholders' perceptions regarding factors that should influence technical debt prioritization?), we ran a set of interviews and focus groups with business and technical stakeholders to discuss their perspectives about business value and the technical debt prioritization rule, during steps 2.1 and 2.2 (cf. Figure~\ref{fig_steps}).

\textbf{1\textsuperscript{st} Goal: Relate the technical debt to its affected value sources to identify its business priority.}
Our objective was to trace the configuration items affected by a technical debt item to its impacted IT Assets and Value Sources. To achieve this goal, we collected (step 1.1) a list of \textit{technical debt items}, (step 1.2) information about \textit{configuration items} and their relationship, (step 1.3) \textit{IT Assets}, and (step 1.4) \textit{Value Sources}. After the initial data collection, we (step 1.5) \textit{tested the technical debt prioritization} with a business stakeholder to have a first evaluation of the relationship between technical debt and value sources.

\textbf{2\textsuperscript{nd} Goal: Classify the business value of value sources and find a consensus regarding business value and technical debt prioritization rule}. As we verified before \cite{icsme2018}, different stakeholders have different opinions about the business perspective which makes finding a consensus on the value source business classification a necessity. 

After achieving the first goal, we collected the (step 2.1) \textit{value source business value classification} from different stakeholders as well as the different perspectives for the (step 2.2) \textit{technical debt prioritization rule}. Our objective was to identify disagreements between stakeholders and promote a consensus about the value and prioritization criteria to be used. 

Since the correctness of the data collection is essential to enable a correct technical debt prioritization, during all steps, at least two participants of the same profile (technical or business) reviewed all collected data. Also, participants could review and update all data regularly using the TDM tool. At the beginning of the case study, we established a research policy where the study participants were the sole responsible for providing and updating the data in the TDM tool.

On the first day, the whole team participated in a training about the main technical debt types and concepts, and the concepts of the proposed approach: the configuration items, IT Assets, value sources, and business impact. The training about technical debt became part of the team's onboarding protocol for new members and was repeated twice during the case study when new members became part of the team. 

\subsection{The TDM Tool} To integrate the research into the team's routine, we developed a tool to collect data, evaluate the model, and support technical debt prioritization. The TDM tool enabled monitoring the team's issue tracking system and manage the technical debt lifecycle and prioritization. Thus, it was possible to conduct the technical debt management with the team's development activities without interfering with their routine. It was also possible to enable and disable the tool's functionalities to execute research activities with the participants.

The development of the tool started in September 2019. Since the model was likely to evolve to address new requirements, the TDM tool underwent parallel development with frequent releases during the case study. It had 11 major and 45 minor updates, with features, enhancements, and bug fixes during the case study (April 2020 to September 2020).

In the following, we detail each step of the process of data collection and analysis.

\subsection{Technical debt list} 

Since the team did not use a structured approach to technical debt management, the initial set of technical debt items (Figure~\ref{fig_steps}, step 1.1) was obtained through a focus group with all squad participants. The participants were asked to discuss all existing technical debt they were aware of in the systems they work with. All debt items identified in this focus group were stored in the TDM tool by the Squad Leader. The discussion and this exercise contributed to strengthening the concept of technical debt in the squad.

After that, the team started using the TDM tool to store all technical debt they identified during their work. Identification of technical debt can occur at any time, by any team member. Each technical debt item is registered using the following information: its name, description, creation date, payment date, type, technical priority, and technical effort. The technical priority is a priority given by a technical leader, used to prioritize the issues in the backlog. The technical effort (high-medium-low) is the evaluation of the effort necessary to pay the technical debt.

 Many technical debt items are identified during planning and problem-solving meetings. Once identified, it is possible to register the technical debt in the TDM tool and also to import any issue marked as technical debt from the issue tracking system. During the case study, 30\% of the technical debt items were registered using the TDM tool, and the other 70\% were imported from the backlog issues classified as technical debt.

\begin{figure}
 \begin{center}
 \caption{TD Types}
 \includegraphics[width=.8\columnwidth]{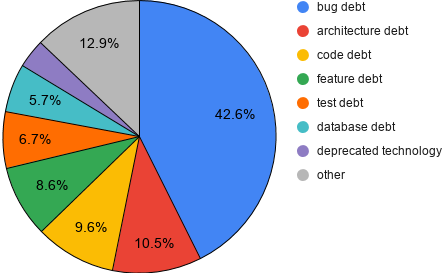}
 \vspace{-.5cm}
 \label{fig_TD_Types}
 \end{center}
 \end{figure}

In our case study, 209 technical debt items were reported and managed in the TDM tool. We started with a total of 140 technical debt items identified from the existing backlog and focus group. Ten of them were paid before the case study start date. During the case study, 69 new TD items were identified, mainly during the sprint execution and evaluation. In the end, a total of 62 TD items were paid, and 137 remained unpaid during the case study period (6 months). The most frequent type of debt was bug debt (42.6 \%), followed by architectural debt (10.5\%), code debt (9.6\%), feature debt (8.6\%), and test debt (6.7\%) (Fig.~\ref{fig_TD_Types}). Bugs were considered technical debt if they were expected to be solved during the sprint period and were postponed for some reason \cite{SEAMAN2011,Codabux2013,LENARDUZZI2021}.

\subsection{Configuration Items}

Since the team did not have updated documentation of its maintained configuration items, we started collecting data about configuration items (Figure~\ref{fig_steps}, step 1.2) through interviews and discussions involving the squad leader and the architect. 
The analyzed data included architecture diagrams, source code structure, modules, services, infrastructure elements, and their related dependencies. All data were stored in the TDM tool, and the existing technical debt items were related to their affected configuration items.

When the participants registered new technical debt items, they could select existing CIs or create new ones if necessary. This process was repeated at each technical debt registration and contributed to the refinement of the configuration item granularity.

We considered the granularity of systems, services, and their main modules to describe the configuration items. A module can be a subset of a system which has a well-defined responsibility. For example, an e-commerce system (composed CI) is composed of authentication and payment modules. It can also be an instance of a system or application. For example, a white-label mobile application (composed CI) is composed of different customized systems of the same application. Note that composed CIs share the code with their components.

\subsection{IT Assets} 

IT Assets are shared between business and technical domains, so the IT Assets were obtained by interviewing the product owner and the squad leader and were reviewed by the CEO (Figure~\ref{fig_steps}, step 1.3). Since the configuration items are from the technical domain, the relationship between IT Assets and their supported configuration items was provided by the squad leader and reviewed by the architect. This case study involves eight IT Assets, with five of them in the \textit{operational} state, two \textit{to-be operational}, and one \textit{legacy}.

\subsection{Value Sources}

The value sources are business domain entities. Therefore, we obtained the first set of value sources and their relationship with IT Assets (Figure~\ref{fig_steps}, step 1.4) from the PO and squad leader, with a review by the CEO.
After that, every technical debt item was linked to existing value sources and---following the same process of the previous entities---if the user missed a value source, it was possible to register a new one. Moreover, this set evolved during the usage of the TDM tool. 

\subsection{Prioritization testing}

After the first value source set collection, we ran a preliminary technical debt prioritization (Figure~\ref{fig_steps}, step 1.5) considering the first evaluation of the value sources provided by the PO. The objective was to test the technical debt prioritization with a controlled scenario, and verify whether we were missing something. We organized three sessions with the PO, where we asked him to classify the technical debt from a set of existing issues from their backlog. For the first set, we randomly selected old issues (20 out of 250 issues from January 1 to December 31, 2019), and in the second set, we selected newer issues (20 out of 123 issues from January 1 to June 30, 2020). Twenty issues were an appropriate number to fit into a one-hour session. We selected old issues for the first session to avoid that his classification would be affected by current business pressures. In other words, if an issue is old, it was not prioritized for a long time and tended to have low priority. This would be a good way to understand the variables that are not addressed by our model. In contrast, in the second session, we selected recent issues to verify whether current business pressure would interfere with the classification. Finally, we ran a third session where we asked him to freely select issues that he thought have a high business priority and should be selected to be paid. He selected 13 issues. After the three sessions, we identified that we must include the ``usage frequency'' variable in the model to improve the value source evaluation. The CEO and the technical leader also confirmed and agreed with the new variable.


\subsection{Value source classification}

The penultimate step to enable the business-driven technical debt prioritization is the classification of the value sources (Figure~\ref{fig_steps}, step 2.1). The stakeholders involved in the decision making must agree on the value source classification to avoid conflicts in technical debt prioritization.

To check the alignment between the participants about how they perceive the business value of their software systems, we asked five different stakeholders (PO, CEO, two developers, and one tester) to categorize 46 value sources as ``core-business'' or not. They also classified the value sources as ``high'' or ``low'' usage regarding their business value and usage frequency. We opted for binary classification to help the stakeholders decide and converge on the classification of what is sufficient for decision making.

During our discussions, we confirmed that the binary business-value classification makes sense. One of the participants, for example, said to convince others, ``there is no medium core business, a feature is core or is not.'' However, the usage frequency can be improved to a range, and also receive the input from a monitoring system. The binary \textit{usage frequency} classification must reflect the relevance of the usage of a value source for decision making. ``High'' means that the usage frequency is relevant from the business perspective, and ``low'' that it is not.

\subsection{Priority rule definition} 
\label{business_focus_group}

To understand the perspectives behind the business prioritization, we evaluated the prioritization rule with a set of five POs, the CEO, and the CSO (Figure~\ref{fig_steps}, step 2.2). To avoid interference between participants, we ran individual interviews and asked them to provide a prioritization rule based on their context (e.g., products, squad, business forces). Table~\ref{table_prioritizations} shows the different prioritizations. 

After the individual interviews, we conducted a focus group with all participants to discuss the different prioritization scenarios provided in the interviews. We first asked the participants who had different perspectives from the majority to discuss their proposal. For example, P6 prioritized value sources that have high usage frequency. After that, we opened the discussion. Since there is a hierarchy between POs, CEO and CSO, the POs talked first, to reduce bias in their opinion. 

\begin{figure}
\begin{center}
\caption{Priority rule considered in the case study}
\includegraphics[width=.6\columnwidth]{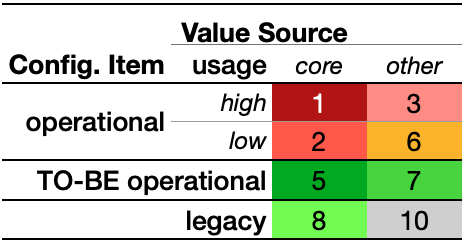}
\vspace{-.5cm}
\label{table_final_priority_rule}
\end{center}
\end{figure}

\section{Impact of the Business-Driven Approach}

\subsection{Answering RQ1: How does the proposed business-driven approach impact technical debt prioritization?}

\begin{table}
\begin{center}
\caption{Technical versus business priorities}
\includegraphics[width=1\columnwidth]{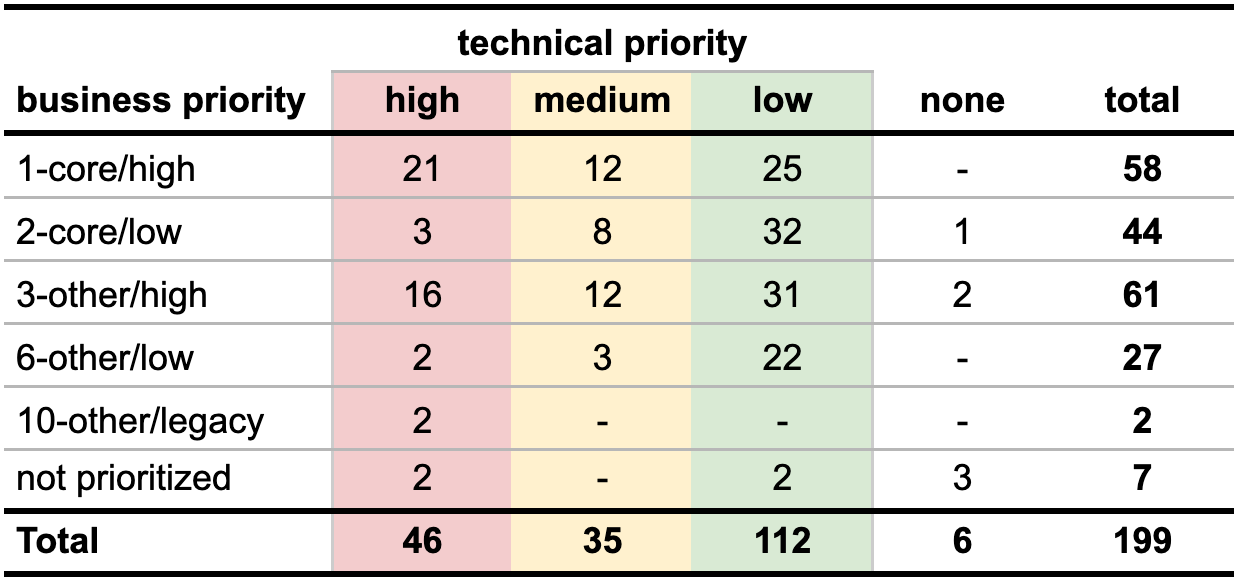}
\vspace{-.5cm}
\label{table_td_busPrio_vs_techPrio}
\end{center}
\end{table}

\begin{table}
\begin{center}
\caption{\% of technical debt payment}
\includegraphics[width=.9\columnwidth]{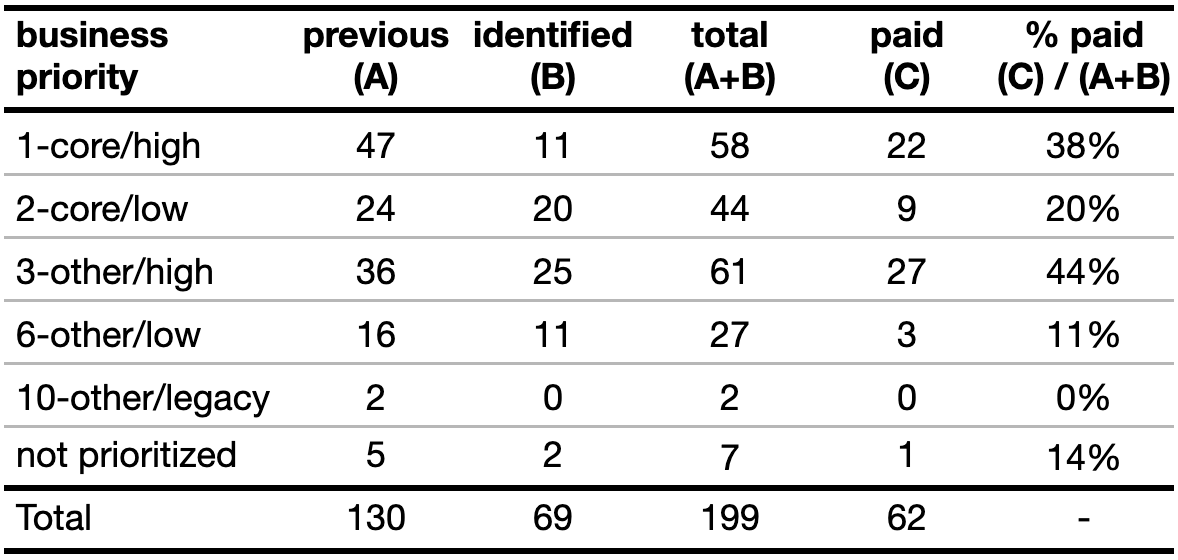}
\vspace{-.5cm}
\label{table_prev_ident_paid}
\end{center}
\end{table}

\noindent\textbf{Technical versus business-driven technical debt prioritization}: Table~\ref{table_td_busPrio_vs_techPrio} shows the tension between the business and technical prioritization. From the 58 items with the highest business classification 1-core/high, only 21 (36\%) also have a high technical priority and would be aligned with business priorities. The discrepancy is higher in the 2-core/low items category since 73\% of the items have the lowest technical priority but may affect core features or business processes. \textbf{The difference between technical and business priorities shows that business prioritization is an additional dimension to support decision-making.}

\begin{figure}
\begin{center}
\caption{Technical debt accumulation trends}
\includegraphics[width=1\columnwidth]{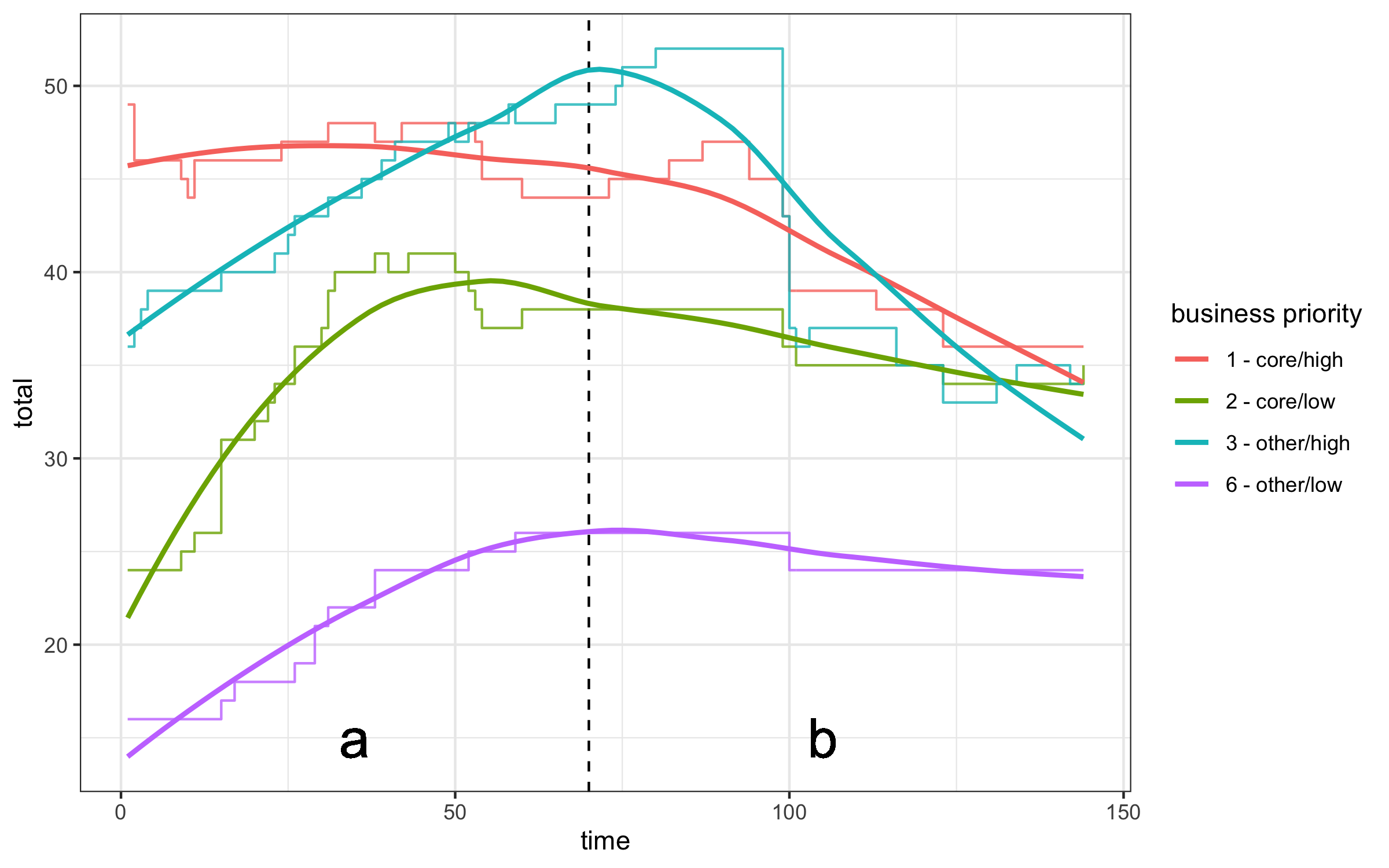}
\vspace{-.5cm}
\label{graph_trends}
\end{center}
\end{figure}

We also analyzed the accumulated series of technical debt during the case study period (143 days). Figure~\ref{graph_trends} presents the graph of the amount of technical debt classified according to its business priority. Each thin line represents the total number of technical debt items identified and paid in a day. For example, if two items are identified, and one is paid, the day has a total of one debt item added to the previous day. The bold lines are the technical debt item accumulation trends. It is important to highlight that we consider the item's identification date, not the registry date. The vertical dotted line indicates when we achieved goals 1 and 2 and started to use our approach to perform business-driven prioritization. The x-axis is divided into periods \textit{a} (technical prioritization) and \textit{b} (business-driven prioritization). 

Table~\ref{table_prev_ident_paid} shows the number of technical debt items registered before the case study, and the number of identified and paid ones during the study. Table~\ref{table_tech_effor} shows the percentage of paid items regarding their technical effort for the most paid items (1, 2, and 3). Our objective is to compare the trends during the period ``b'' and evaluate the forces behind selecting which item should be paid. The amount of paid technical debt classified as \textit{1-core/high} (38\%)  and \textit{3-other/high} (44\%) was higher than the debt classified as \textit{2-core/low} (20\%) and \textit{6-other/low} (11\%). When we consider the paid technical debt's technical effort, we observe that the highest effort was dedicated to the items with the highest business priority (cf. Table~\ref{table_tech_effor}).  22.7\% of the highest business priority items had a high technical effort, while only 11.5\% of the ones classified with business priority 3 had high technical effort. Among the paid items with the highest priority, 5 (22\%) of them were of the ``architectural debt'' type, and no architectural debt with business priorities 2 or 3 was paid. \textbf{Despite the number of paid items with business priority 3-other/high being higher than the ones classified as 1-core/high, the team dedicated more technical effort to pay the technical debt with business priority 1-core/high.}

\begin{table}
\begin{center}
\caption{Technical effort of the paid technical debt}
\includegraphics[width=.7\columnwidth]{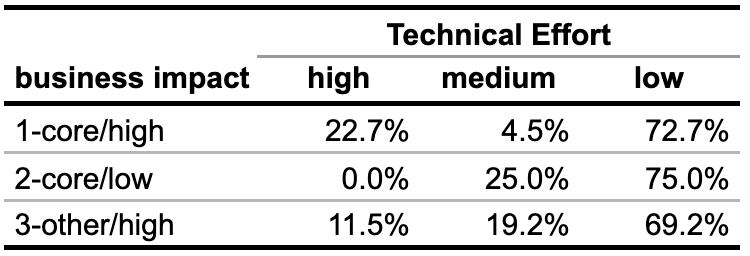}
\vspace{-.5cm}
\label{table_tech_effor}
\end{center}
\end{table}

\subsection{Answering RQ2: What are the business stakeholder's perceptions regarding factors that should influence technical debt prioritization?}

During the meetings to discuss the value source classification and the prioritization rule, the different perspectives were frequently apparent. It was possible to identify some factors that are behind what the participants considered in their arguments. 

Below we discuss the scenarios of conflicts during the meetings of the value source classification and the priority rule definition. It is important to highlight that while we bring people from different projects together, it is expected that they have different opinions about the topic under analysis.

\begin{table}
\begin{center}
\caption{Agreement on value source classification}
\includegraphics[width=.6\columnwidth]{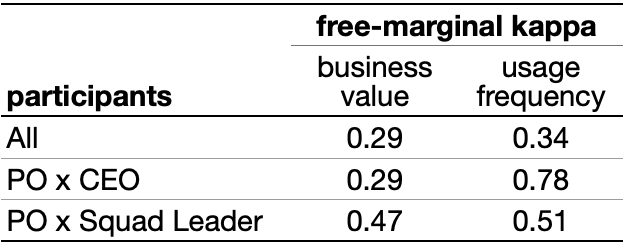}
\vspace{-.5cm}
\label{table_VS_agreement}
\end{center}
\end{table}

\noindent\textbf{Conflict analysis of the value source classification:} Table \ref{table_VS_agreement} shows the Kappa \cite{kappa} agreement among the participants, regarding their opinion about the business-value and usage frequency of the value sources. First, with all participants (45 cases, two categories, and five raters), we had 0.29 Kappa agreement on the business value and 0.34 Kappa agreement on the usage frequency. This shows how the perception of what is core-business or not is different among the team members. We observed the same level of agreement (0.29) on the business value and 0.78 Kappa agreement regarding the usage frequency between the business stakeholders (PO and CEO). 

The disagreement between the CEO and PO was a result of a different evaluation dimension about core business from other study participants. The CEO's perception about the business value is customer and marketing-centric whereas the PO has a more functional-centric perception. For example, the CEO evaluated a report as core-business while the PO did not---for the CEO, the report ``is on my sales presentations'' but the PO considered that the report was a side-feature and was not related to what the product was supposed to do as core-business. The PO's classification was more aligned with the definition of a core business feature, but the report had enough business value to maintain its classification as a core business. Both agreed that the report had low usage. 

Finally, the PO and Squad leader had a higher agreement for the two variables---0.47 on the business value classification and 0.51 on the usage frequency. Unlike the PO-CEO case, the perception of business value was aligned between the PO and the squad leader. Both share a routine of feature-driven decision-making, making them share the same level of perception of how core-business features and processes are. Their disagreement was mostly related to a lack of understanding of how features work and how processes affect customer routines and outdated information about business contracts and feature usage. 

\noindent\textbf{Incident versus Technical debt prioritization:} Both technical and business stakeholders often evaluated a value source or technical debt priority considering an incident scenario. Many times it was necessary to highlight that technical debt and incident prioritization are different. For example, the authentication feature of all products and services was, at a first evaluation, classified as ``core-business'' because ``if the authentication is not working, the user cannot use the system.'' A way to expose the core value sources was to ask, ``What are the customers paying for?''.  For example, in an e-commerce system, the customer does not pay for the system to authenticate users. Although essential for the system usage, the customer pays for selling the product, a core-business feature. On the other hand, if the authentication causes service disruption, the incident must be prioritized and probably will gain a high priority. Note that technical debt can cause incidents, and maybe debt should be prioritized if that is the case, but as a result of a different decision-making process.

The ``authentication'' feature example came up in one of our meetings. One of the participants realized that, besides proposing the authentication as a core-business, they did not prioritize an architectural debt item on their authentication solution for almost one year. Other technical debt items always gained priority and were paid before the authentication one. They said, ``We still can wait for a while to pay it.''

Another way to identify the core business features and processes was discussed by one of the participants. While convincing others that a feature was core-business, he asked, ``Does the feature have any business rule on it? It is only a CRUD!''. It was a good point to separate features that are only data management (for example, CRUDs for some entities) from other features that aggregate value and have more business rules, like features that process sales transactions.

\label{sec_percep_prior}

\begin{table}
\begin{center}
\caption{Prioritization rules proposed by business stakeholders}
\includegraphics[width=.9\columnwidth]{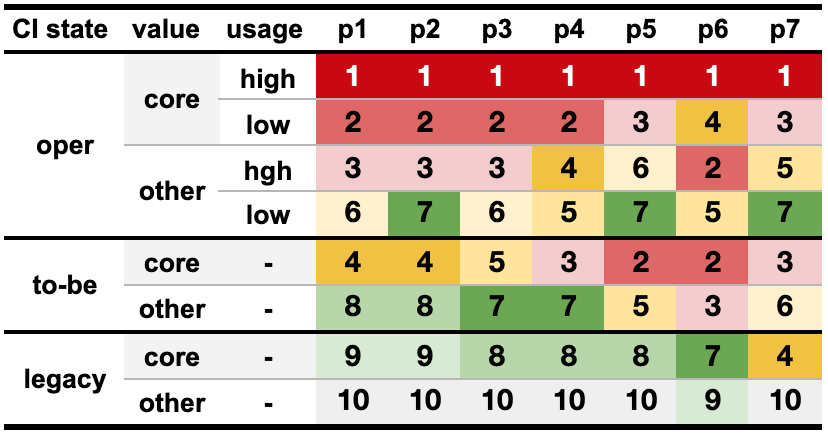}
\vspace{-.5cm}
\label{table_prioritizations}
\end{center}
\end{table} 

\noindent\textbf{The perception of risk affected the technical debt prioritization rule:} Table~\ref{table_prioritizations} presents the different prioritization rules suggested by the participants after the focus group (Section \ref{business_focus_group}). As we can see, there are two unanimous prioritizations: all participants consider that technical debt, which affects operational IT assets and core/high value sources, must receive the highest priority, while the ones that affect legacy IT Assets must receive the lowest priorities. More than one participant declared that \textit{``Technical debt which affects operational core-business commonly used features and processes must be prioritized.''}. 

We also analyzed which categories received the highest classifications (1 to 3) (Table~\ref{table_temp}). High usage was the most prioritized, with 78.6\%, followed by 60.7\% of both core and operational. To-be received 64.3\% medium and 50.0\% of low prioritizations. Legacy was the lowest priority, with 92.0\% of lower classifications, followed by to-be and other with 50.0\%. The \textit{high usage} prevalence is aligned with existing findings \cite{featureValue} that ``in most of the cases the higher the usage, the higher the perceived value of a feature''.

\begin{table}
\begin{center}
\caption{Percentage of the decomposed variables considered in the proposed technical debt prioritization rules}
\includegraphics[width=.6\columnwidth]{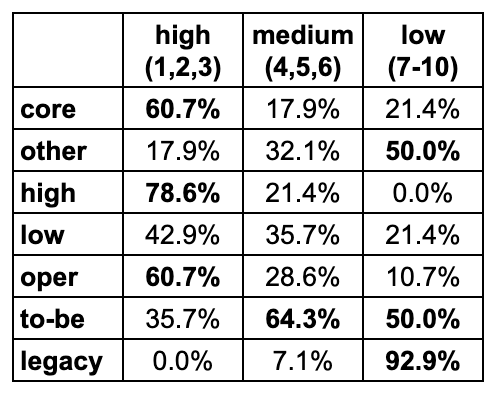}
\vspace{-.9cm}
\label{table_temp}
\end{center}
\end{table}

We refer to our participants using the identifiers P1 to P7. Participants P1, P2, and P3 provided the same prioritization with different perceptions about what is \textit{high}, \textit{medium} or \textit{low}. For example, P2 considers \textit{other/low} as a \textit{low} priority (7) while P1 considers it a \textit{medium} priority (6). They prioritize \textit{core/high} while \textit{other/legacy} had the lowest priority. 

P5 was the only one who prioritized \textit{core-legacy}. His context involves many solutions that must be certified, and the process of feature and system replacement has a delay due to certification procedures that involve third-party certification companies. He said: \textit{``Since we cannot release features as frequent as the other squads, we have to prioritize technical debt on \textit{to-be} configuration items. We cannot forget the existing debt of the systems and modules that will die (\textit{legacy}) since they are slow to die.''}

The perception of risk of a technical debt item attracted different perspectives. For example, participants P1, P2 and P3 classified the priority of \textit{core/to-be} as \textit{medium} (4 and 5) while the others (P4 to P7) prioritized it as \textit{high} (2 and 3). When asked to explain their motivations, P1 said that \textit{``since the new feature or system was not delivered, we have more opportunity to handle the technical debt, negotiate and delay it, different from the case where the debt is already creating risk on an operational feature or system."}

On the other hand, participant P5 argued that \textit{``the risk of debt on an operational feature is already known, we already decided before it went on-line. We know how many users can be affected by it and if it is causing incidents. On the other hand, the debt on a to-be feature is unknown. Will the user face incidents due to the debt? Will we have time and the ability to pay it in the future?''}  Another argument in the direction of prioritizing technical debt that affects a \textit{to-be} CI came from the perception that \textit{``every debt we pay on a feature or system that is not yet operational (to-be), will be on an operational system in the future.''} If the teams do not prioritize debt on to-be features and systems, they will have to prioritize it in the future, when it becomes operational.

\begin{table}
\begin{center}
\caption{ Factors that affected the TD prioritization,from the business perspective }
\includegraphics[width=1\columnwidth]{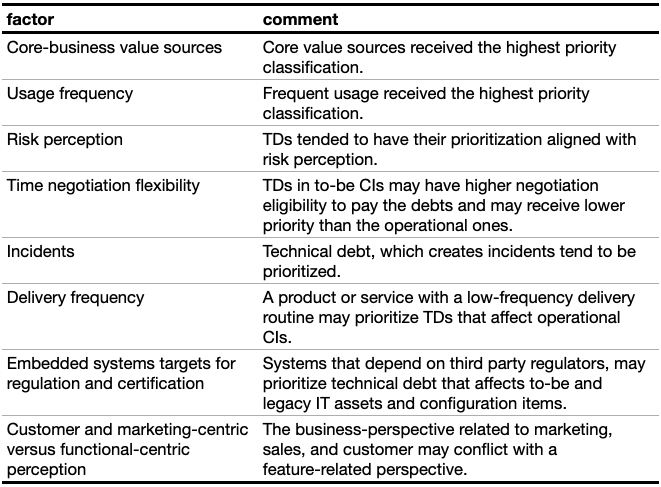}
\vspace{-.9cm}
\label{table_factors-rq2}
\end{center}
\end{table}

\textbf{After analyzing the participants' arguments, we identified eight factors that influenced the technical prioritization}, considering the business perspective. Table \ref{table_factors-rq2} summarizes them. In this case study, it was unanimous that if a technical debt item affects a \textit{core-business} feature or a feature that is heavily used by users (\textit{usage frequency}), it should have a high priority. The risk perception related to the unknown impact of technical debt also affected the prioritization. A high-risk perception implied a higher TD prioritization. 

The \textit{flexibility to negotiate the time} to delivery also affected decision making. Situations where there is flexibility to negotiate and expand the delivery time tended to lower the technical debt priority. As one participant said: ``If I have to choose between paying a TD in operation and one still in the development cycle, I choose the operation one. It is feasible to negotiate scope and time under development; in production, I cannot. The debt is already creating interest.'' Participants also agreed that if a TD item is creating \textit{incidents}, then its payment must be prioritized. The \textit{frequency in which teams deliver} releases in production also affected the prioritization decision making. It is related to the ``time negotiation flexibility'' factor. Teams with low-frequency delivery have more time to deal with TD in the development cycle and prioritize TD in production (operational). The PO of a team that deals with \textit{certified embedded systems} explained that the cost to change the system to pay a TD item in production is so high that they must prioritize all known TD in the development phase before being delivered. They also deal with many devices using legacy systems, which makes them prioritize technical debt in legacy systems. 

Finally, we observed misalignment between the arguments that drive technical debt prioritization, according to different business-level participants. Product owners tend to have a more \textit{feature-related perspective}, their motivations to define the prioritization were more related to deadlines, milestones, and scope. On the other hand, the directors and account managers tend to have arguments related to \textit{marketing and sales perspectives}. Complementary perspectives should be aligned to support the overall technical debt management decision making.

\section{Applying the approach}

To apply the approach for business-driven technical debt management, we suggest to start with one team and ensure the participation of at least one technical stakeholder and one business stakeholder. An initial training with the team to reach common ground on what is technical debt builds the foundation for technical debt management and can be followed by a workshop with a technical leader and a business stakeholder to identify the initial set of IT Assets that will be the scope of the technical debt management. For each IT Asset, the initial set of value sources need to be identified and classified according to their business value (core/other) and usage frequency (high/low). To identify the initial set of technical debt items to be managed and to define the first prioritization rule, a workshop with the technical team is suitable. An iterative and incremental process should then be followed:

\begin{enumerate}
  \item for each technical debt item, identify and register the affected configuration item. We suggest beginning with a two-level granularity, e.g., system/service and module;
  \item relate the configuration item with one of the IT Assets;
  \item relate the technical debt with one of the selected IT asset's value sources;
  \item review the configuration items, IT Assets and value sources;
\end{enumerate}

\section{Threats to Validity}

\textit{Internal validity}: During the whole study, we had to avoid hierarchy bias since we were dealing with conflict situations between different knowledge levels, e.g., business versus technical stakeholders, and different hierarchy levels, e.g., directors and product owners. To overcome this bias, we used an approach to collect information individually before opening it to discussion and letting participants at lower levels of the hierarchy speak first.  

\textit{External validity}: The presented results are related to one company and one team. They cannot be considered generalizable. However, the evaluated approach's building blocks do not use concepts or rules particular to the company. The business factors that affect the TD prioritization resulted from the seven business participants' perceptions and we cannot claim generalizability to other participants.

\section{Related Work}

The research field currently lacks business-related criteria for decision making and approaches for technical debt prioritization. A recent systematic literature review on technical debt prioritization \cite{SLR-TD-prioritization} reveals the scarcity of approaches that account for cost, value, and resource constraints as well as a lack of industry evaluation. A systematic mapping study~\cite{Ribeiro:2016} identified 14 decision making criteria that can be used by development teams to prioritize the payment of TD items. The identified studies concentrated on two types of debt (defect and design), and the only studies that consider business-related criteria are concentrated on ``cost-benefit''. Our work is positioned to contribute to filling this gap.

Ribeiro et al.~\cite{ribeiro2017} present a strategy for TD management that uses multiple decision criteria to decide when to pay debt items off. Their work proposes a configurable multi-criteria decision approach based on weights assigned to 14 categories. Some of the criteria can be driven by business forces, like the customer, severity, and cost-benefit, but the approach considers the classification for each technical debt item individually, done by a software engineer. Their approach is different from ours since they do not consider the IT artifacts affected by the technical debt or its business value. We also work with the definition of a general prioritization rule applied to technical debt, despite the individual technical debt evaluation.

Our work is aligned with Martini and Bosch~\cite{martini-boch-2015} who provide nine prioritization aspects for architectural technical debt, identified by business and technical participants. Their identified aspects (e.g., competitive advantage, specific customer value, market attractiveness) can be used to guide the classification of value sources. They also identified different conflicts regarding the prioritization between the POs and the software architects.

\section{Conclusions}

We performed an industrial case study to evaluate how a business prioritization approach for technical debt works in a real scenario. 

We observed misalignment regarding the prioritization of technical debt, the value source classification, and the prioritization rule. These conflicts are expected when stakeholders of different domains are involved. Our business-driven approach contributes to the alignment of the business perspectives for technical debt prioritization.

We applied the proposed approach supported by a five-months case study where we associated technical debt items with business-value elements. We also identified and solved conflicting business perspectives among stakeholders. We observed a downward trend in the resolution of technical debt items that are related to high business priority. We also found that much goes on behind the scenes regarding the prioritization decision making. Finally, after a set of interviews and focus groups, we identified eight business factors that affect decision making regarding technical debt.

As future work, we plan to develop guidelines to support the business-driven decisions regarding technical debt prioritization. To measure the effort of applying the approach and tool, we plan to apply them in another team and company.

\bibliographystyle{IEEEtran}
\bibliography{bibliography/techdebt}

\end{document}